\documentclass[a4paper]{jpconf}
\usepackage{graphicx}

\makeatletter
\def\slash#1{{\mathpalette\c@ncel{#1}}} 
\makeatother

\newcommand\beq{\begin{eqnarray}}
\newcommand\eeq{\end{eqnarray}}

\newcommand\la{\langle}
\newcommand\ra{\rangle}

\def\xhat{\widehat{x}}
\def\zhat{\widehat{z}}

\begin{document}
\title{On the contribution of twist-3 multi-gluon correlation functions 
to single transverse-spin asymmetry in SIDIS}

\author{Hiroo Beppu$^1$, Yuji Koike$^2$, Kazuhiro Tanaka$^3$ and Shinsuke Yoshida$^1$}

\address{$^1$ Graduate School of Science and Technology, Niigata University,
Ikarashi, Niigata 950-2181, Japan}
\address{$^2$ Department of Physics, Niigata University,
Ikarashi, Niigata 950-2181, Japan}
\address{$^3$ Department of Physics, 
Juntendo University, Inzai, Chiba 270-1695, Japan}

\ead{tanakak@sakura.juntendo.ac.jp}

\begin{abstract}
We study the single spin asymmetry (SSA)
induced by purely gluonic correlation inside a nucleon, 
in particular, by the three-gluon correlation functions in the transversely polarized nucleon, $p^\uparrow$. 
This contribution is embodied as a twist-3 mechanism 
in the collinear factorization framework 
and controls the SSA to be observed in the $D$-meson production 
with large transverse-momentum 
in semi-inclusive DIS (SIDIS),
$ep^\uparrow \rightarrow eDX$. 
We 
define 
the relevant three-gluon correlation functions in the nucleon, and determine their complete set 
at the twsit-3 level taking into account 
symmetry constraints in QCD. 
We derive the single-spin-dependent cross section for the $D$-meson production in SIDIS,
taking into account all the relevant contributions at the twist-3 level.
The result 
is obtained in a manifestly gauge-invariant form as the factorization formula in terms of 
the three-gluon correlation functions 
and reveals the five independent structures with respect to the dependence on the azimuthal angle 
for
the produced $D$ meson. 
We also demonstrate the remarkable relation between the twist-3 single-spin-dependent cross section 
and twist-2 cross sections for the $D$-meson production, 
as a manifestation of universal structure 
behind 
the SSA in a variety of hard processes.
\end{abstract}

\section{Introduction}
\begin{figure}[h]
\begin{minipage}{24pc}
\begin{center}
\includegraphics[width=12pc]{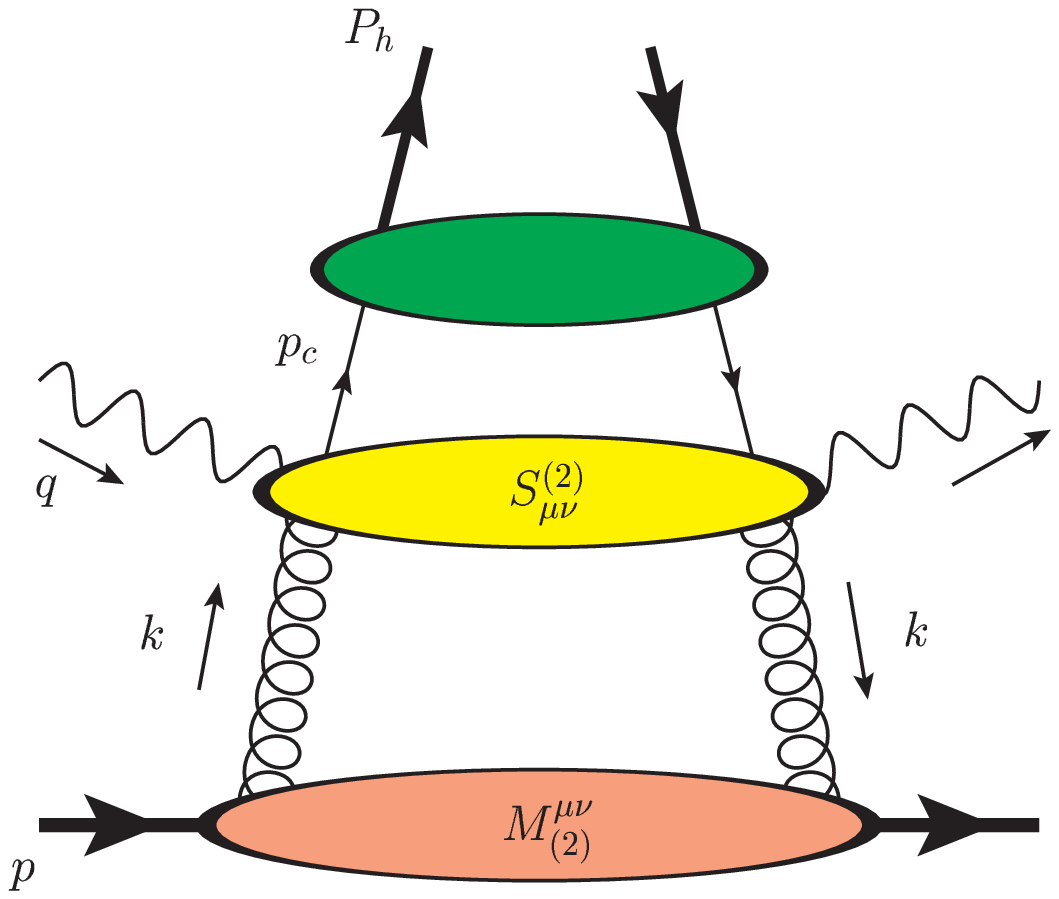}~
\includegraphics[width=12pc]{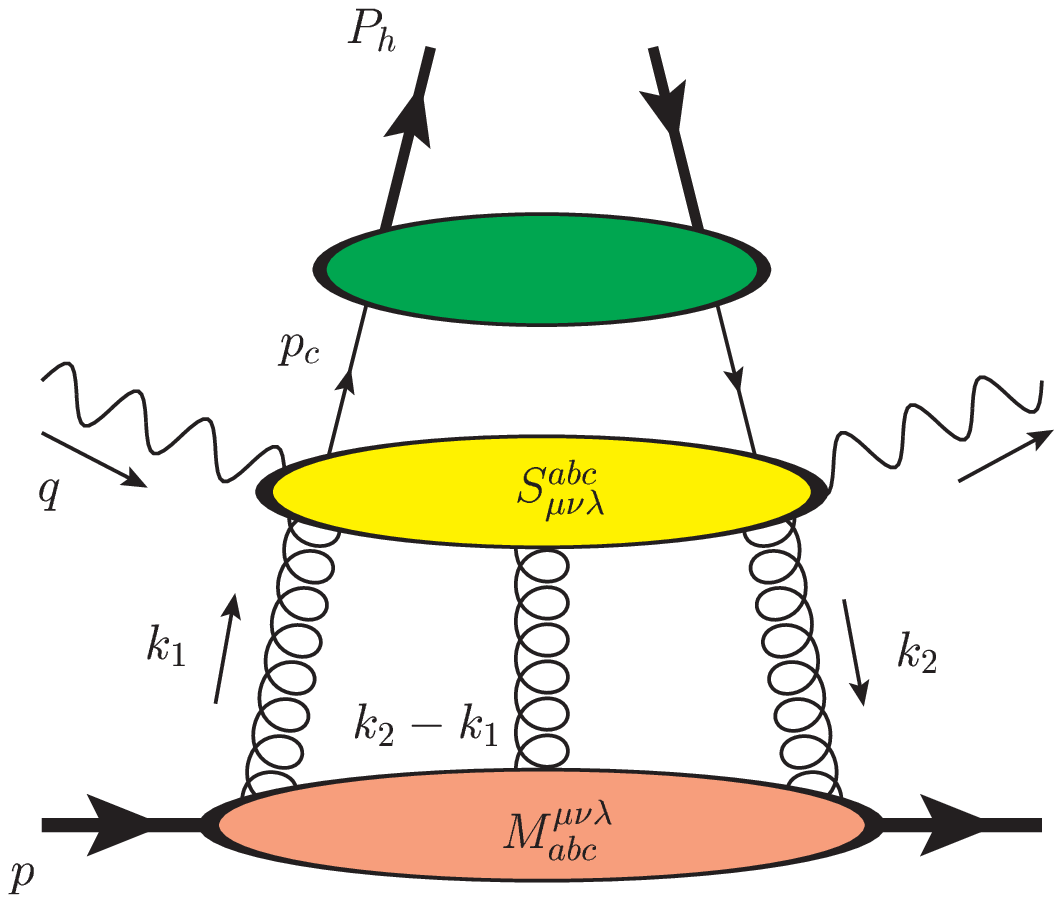}
\end{center}
\vspace{-0.2cm}
\hspace{2.2cm}(a)\hspace{4.8cm}(b)
\end{minipage}\hspace{2pc}%
\begin{minipage}{12pc}\caption{
Generic diagrams for 
hadronic tensor of $ep^\uparrow\to eDX$
induced by gluonic effects in the nucleon.  Each one is
decomposed into nucleon matrix element (lower blob), $D$-meson matrix element (upper blob),
and partonic hard scattering part by the virtual photon through the photon-gluon fusion (middle blob). 
\label{fig1}
}
\end{minipage}
\end{figure}

The single transverse-spin asymmetry (SSA) in the semi-inclusive DIS (SIDIS), 
$e(\ell)+p(p,S_\perp)\to e(\ell')+h(P_h)+X$,
arises as {\em T-odd} effect 
in the cross section for the scattering of transversely polarized nucleon with momentum $p$
and spin $S_\perp$,
off unpolarized lepton
with momentum $\ell$, 
observing a hadron $h$ with momentum $P_h$ 
in the final state. Here, $q=\ell-\ell'$, $Q^2=-q^2$,
and $Q \gg \Lambda_{\rm QCD}$.
The SSA can be observed also in $pp$ collisions,
the pion production $p^\uparrow p\to\pi X$~\cite{To},
and the Drell-Yan and direct-$\gamma$ production, 
$p^\uparrow p\to \gamma^{(*)} X$~\cite{JQVY06DY}.
Similarly as 
these 
examples, 
the SSA in the SIDIS requires,
(i) nonzero transverse-momentum $P_{h\perp}$ originating 
from transverse motion
of quark or gluon; 
(ii) nucleon helicity flip in the cut diagrams for the cross section, corresponding
to the transverse polarization $S_\perp$; 
and (iii) interaction beyond Born level to produce the
interfering phase between the LHS and the RHS of the cut in those diagrams. 
When $P_{h\perp}\ll Q$, all (i)-(iii) may be generated nonperturbatively
from 
the T-odd, transverse-momentum-dependent (TMD) parton distribution/fragmentation
functions 
such as the Sivers function (see~\cite{Bacchetta:2006tn,JQVY06SIDIS}).
By contrast,
for large $P_{h\perp}\gg \Lambda_{\rm QCD}$,
(i) should come from perturbative mechanism
as the recoil from the hard (unobserved) final-state partons, 
while nonperturbative effects 
can participate in the other two, (ii) and (iii), allowing us to obtain large SSA.
This 
is realized 
with the twist-3 distribution/fragmentation functions
in 
the collinear-factorization framework.
In \cite{EKT07}, we 
derived the corresponding factorization formula,
in the leading-order 
(LO) 
perturbative QCD,
for the SSA
associated with the twist-3 
quark-gluon correlation inside the nucleon,
and provided a practical procedure to calculate the 
relevant partonic hard parts
manifesting their gauge invariance at the twist-3 level.
A similar twist-3 mechanism for the SSA can also be generated by the purely gluonic correlations
inside the nucleon.
In particular, the SSA to be observed in the production of the
$D$ meson with large transverse-momentum $P_{h\perp}$
is controlled by the corresponding gluonic
contributions, shown in figure~\ref{fig1},
through the photon-gluon fusion to create a pair
of 
$c$ and $\bar{c}$ quarks, one of which fragments into the $D$ meson.
In this report we discuss those gluonic effects, 
taking into account the nonzero masses, $m_c$ and $m_h$, for the charm quark and the $D$ meson~\cite{bkty};
our SSA formula 
for the high $P_{h\perp}$ $D$-meson production, $ep^\uparrow\to eDX$, 
differs from the corresponding result in the literature~\cite{KQ08}, and
the origin of the discrepancy will be clarified.  

\section{Three-gluon correlation functions in the transversely polarized nucleon}
Figure~\ref{fig1}(a) gives rise to the twist-2 contribution for the unpolarized 
SIDIS, $ep \to eDX$, and the corresponding cross section is expressed by the factorization formula
associated with the unpolarized gluon-density distribution for the nucleon,
\begin{equation}
G(x)
=-\frac{g_{\beta \alpha}}{x}\int{d\lambda\over 2\pi}e^{i\lambda x}
\langle p|F_a^{\beta n}(0)F_a^{\alpha n}(\lambda n)
|p\rangle, 
\label{gx}
\end{equation}
as the nucleon matrix element of
the gauge-invariant lightcone bilocal operator
using the gluon field-strength tensors, $F_a^{\alpha n} (\lambda n)\equiv F_a^{\alpha \beta}(\lambda n)n_{\beta}$,
$F_a^{\alpha\beta}=\partial^\alpha A^\beta_a
-\partial^\beta A^\alpha_a +gf_{abc}A_b^\alpha A_c^\beta$;
here, $p^\mu=(p^+,0,\mathbf{0}_\perp)$ is the nucleon momentum
regarding as lightlike 
up to the twist-3 accuracy,
$n^\mu=(0,n^-, \mathbf{0}_\perp)$ is another lightlike vector satisfying $p\cdot n=1$, and
we have 
suppressed 
the gauge-link operator
which connects 
the field strength 
tensors 
so as to ensure
the gauge invariance.
Likewise, the ``three-gluon distribution'' functions, relevant to the 
SSA arising 
from figure~\ref{fig1}(b), are defined through
the gauge-invariant correlation function of the three field-strength tensors,
\beq
{\cal M}^{\alpha\beta\gamma}_{F,abc}(x_1,x_2)
&&
\!\!\!\!\!\!\!\!
=
-gi^3\int{d\lambda\over 2\pi}\int{d\mu\over 2\pi}e^{i\lambda x_1}
e^{i\mu(x_2-x_1)}
\la pS|F_b^{\beta n}(0)F_c^{\gamma n}(\mu n)F_a^{\alpha n}(\lambda n)
|pS\ra
\nonumber\\
&&\!\!\!\!\!\!\!\!
=\frac{3}{40}d^{abc} O^{\alpha\beta\gamma}(x_1,x_2)
-\frac{i}{24}f^{abc}N^{\alpha\beta\gamma}(x_1,x_2) ,
\label{MFabc}
\eeq
for the nucleon with
the transverse spin vector $S^\mu =(0,0, \mathbf{S}_\perp)$,
normalized as $S^2=-\mathbf{S}_\perp^2=-1$.
This may be regarded as an extension of the 
quark-gluon
correlation functions discussed in, e.g., \cite{EKT07}, 
with $x_{1,2} \equiv k_{1,2}\cdot n$ denoting the relevant momentum fractions of the gluons
shown in figure~\ref{fig1}(b). Here, $d^{abc}$ and $f^{abc}$ are the usual symmetric
and anti-symmetric structure constants of the color SU(3) group, and
the constraints from 
hermiticity, 
invariance 
under the parity and time-reversal transformations,
and the permutation symmetry among the participating gluon fields allow the further decomposition
in terms of the real, Lorentz-scalar functions, $O(x_1, x_2)$ 
and $N(x_1, x_2)$, which are both symmetric under the interchange $x_1 \leftrightarrow x_2$, such that~\cite{bkty}
\beq
&&
\hspace{-0.9cm}
O^{\alpha\beta\gamma}(x_1,x_2)\nonumber\\
&&
\!\!\!\!\!\!
=2iM_N\left[
O(x_1,x_2)g^{\alpha\beta}\epsilon^{\gamma pnS}
+O(-x_2,x_1-x_2)g^{\beta\gamma}\epsilon^{\alpha pnS}
+O( x_2-x_1, -x_1)g^{\gamma\alpha}\epsilon^{\beta pnS}\right],
\label{3gluonO}
\eeq
and similarly for $N^{\alpha\beta\gamma}(x_1,x_2)$,
at the twist-3 accuracy; here, $\epsilon^{\gamma pnS}\equiv \epsilon^{\gamma \lambda \mu \nu}p_\lambda n_\mu S_\nu$. 
$O(x_1, x_2)$ and $N(x_1, x_2)$ are defined as dimensionless, 
accompanying the nucleon mass $M_N$ in (\ref{3gluonO}),
and are associated, respectively, with the $C$-odd and $C$-even combinations of the three gluon operators
in (\ref{MFabc}), satisfying
$O(x_1,x_2)=O(-x_1,-x_2)$ and
$N(x_1,x_2)=-N(-x_1,-x_2)$.
Indeed, $O(x_1,x_2)$ and $N(x_1,x_2)$ 
constitute a complete set 
for expressing 
gluonic correlation effects inside the nucleon 
at the twist-3 level.
We note that the other two functions, in addition to $O(x_1,x_2)$ and $N(x_1,x_2)$, 
were introduced in \cite{Ji92} to constitute 
a basis of three-gluon functions at the twist-3 level,
but this basis proves to be redundant, i.e., the two functions $O(x_1,x_2)$ and $N(x_1,x_2)$ 
form a genuine complete set~\cite{bkty}. 
The authors of \cite{Braun09} also pointed out that there
are only two independent pure gluonic functions at twist-3,
in the study of the 
evolution equations for the twist-3 distributions.

\section{Collinear expansion and gauge invariance 
for the three-gluon correlation effects}
We calculate the hadronic tensor $W^{\mu\nu}(p,q,P_h)$, represented by figure~\ref{fig1}.
We are interested in the contribution 
in which the $c$ and $\bar{c}$ quarks are created through the photon-gluon
fusion process and one of them fragments into a $D$ 
($\bar{D}$)
meson, so that
the corresponding fragmentation function $D(z)$, with $z$ being the relevant momentum fraction,
is
factorized from $W_{\mu\nu}$ as
\beq
W_{\mu\nu}(p,q,P_h)=
\int{dz\over z^2}D(z) w_{\mu\nu}\left(p,q,p_c\right) , 
\label{Wmunu}
\eeq
where the summation over the $c$ and $\bar{c}$ quark contributions is implicit,
and $p_c$ is the momentum of the $c$ (or $\bar{c}$) quark with mass $m_c$;
using
the lightlike vector $w$ of $O(1/Q)$ satisfying $P_h \cdot w=1$,  
we have, $p_c^\mu = P_h^\mu / z + rw^\mu$,
with $r= ( m_c^2z - m_h^2/z )/2$ to satisfy $p_c^2=m_c^2$, and, 
at the leading twist-2 accuracy for the quark-fragmentation process,
we set $w^\mu=p^\mu/(P_h\cdot p)$.
We also note that,
in the LO with respect to the QCD coupling constant for the partonic hard scattering parts,
the contribution of figure~\ref{fig1}(a) can not give rise to the SSA,
due to the symmetry properties
of the correlation functions of two 
gluon fields in the polarized nucleon (see Appendix~A in \cite{bkty}).  
So we shall focus on the analysis of figure~\ref{fig1}(b).
We work in Feynman gauge and 
apply the collinear expansion to the hard scattering part in figure~\ref{fig1}(b),  
keeping all the terms contributing up to the relevant accuracy of twist-3.   
We perform the corresponding calculation on the basis of generalization 
of our previous formulation~\cite{EKT07} that was developed for the case of the SSA 
for the pion production induced by the quark-gluon correlation inside the nucleon. 
For simplicity,
we shall henceforth write $w_{\mu\nu}\left(p,q,p_c\right)$ of (\ref{Wmunu}) as
$w\left(p,q,p_c\right)$,
omitting the indices 
for 
virtual photon.

The contribution from figure~\ref{fig1}(b)
to $w(p,q,p_c)$ can be written as
\beq
w\left(p,q,p_c\right)=
\int{d^4k_1\over (2\pi)^4}\int{d^4k_2\over (2\pi)^4}
\,S_{\mu\nu\lambda}^{abc}\left(k_1,k_2,q,p_c\right) 
M^{\mu\nu\lambda}_{abc}(k_1,k_2),
\label{wtensor}
\eeq
where $S_{\mu\nu\lambda}^{abc}(k_1,k_2,q,p_c)$ is the partonic hard scattering
part (middle blob of figure~\ref{fig1}(b)), and 
\beq
M^{\mu\nu\lambda}_{abc}(k_1,k_2)=g \int\, d^4\xi\int\, d^4\eta\, e^{ik_1\xi}e^{i(k_2-k_1)\eta}
\la pS | A_b^\nu(0)A_c^\lambda(\eta)A_a^\mu(\xi) |pS\ra
\label{Mk1k20}
\eeq
is the corresponding nucleon matrix element (lower blob),
%
including one QCD coupling constant~$g$.
In (\ref{wtensor}), a real contribution relevant to the spin-dependent cross section
occurs from an imaginary part of the color-projected hard part
$C_\pm^{bca} S^{abc}_{\mu\nu\lambda}(k_1,k_2,q,p_c)$ with 
$C_+^{bca}=if^{bca}$ and
$C_-^{bca}= d^{bca}$,
since $C_\pm^{bca} M^{\mu\nu\lambda}_{abc}(k_1,k_2)$ 
are pure imaginary quantities~\cite{bkty}.
This means that only the pole contribution produced by an
internal propagator in the hard part 
can give rise to SSA.

\begin{figure}[h]
\includegraphics[width=18pc]{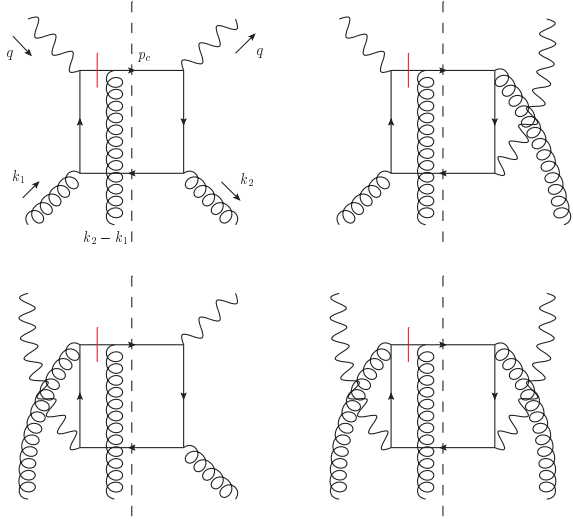}\hspace{2pc}%
\begin{minipage}[b]{17.8pc}
\caption{\label{fig2}Feynman diagrams for the partonic hard part in figure~\ref{fig1}(b), representing the 
photon-gluon fusion subprocesses that give rise to the ``surviving'' pole contribution for $ep^\uparrow\to eDX$
in the LO with respect to the QCD coupling constant. 
The short bar on the internal $c$-quark line
indicates that the pole part is to be taken from that propagator.  
In the text, momenta are assigned as shown in the upper-left diagram, where
$p_c$ denotes the momentum of the $c$-quark fragmenting into the $D$-meson
in the final state.  The mirror diagrams also contribute.}
\end{minipage}
\end{figure}
In the LO with respect to the QCD coupling constant, we find that 
the diagrams shown in figure~\ref{fig2}, together with their
mirror diagrams, give rise to the ``surviving'' pole contributions; 
here, 
the quark propagator with a short bar 
produces the corresponding pole contribution.
The other pole contributions
turn out to cancel among themselves
after summing the contributions of all the leading-order diagrams
for (\ref{wtensor}).  
With the assignment of the momenta $k_1$ and $k_2$ of gluons as shown in figure~\ref{fig2}, 
the condition for those poles is given by $(p_c -k_2 +k_1)^2-m_c^2=0$.
After we perform the collinear expansion 
and reach the collinear limit, $k_i\to x_ip$ ($i=1,\ 2$), with $x_1, x_2$
and $x_2-x_1$
representing the longitudinal momentum fractions of the relevant three gluons,
this condition reduces to $x_1=x_2$
and hence
a pole of such type is referred to as the soft-gluon pole (SGP).  
In the following, we assume that $S_{\mu\nu\lambda}^{abc}\left(k_1,k_2,q,p_c\right)$
in (\ref{wtensor}) represents the sum of the 
contributions of the diagrams in figure~\ref{fig2} and their mirror diagrams,
in which 
the barred propagator 
is replaced by its pole contribution.
We write the hard part 
$S_{\mu\nu\lambda}^{abc}\left(k_1,k_2,q,p_c\right)$ 
as $S_{\mu\nu\lambda}\left(k_1,k_2\right)$, for simplicity,
suppressing the color indices $a,b,c$
and the momenta $q$ and $p_c$, and,
correspondingly, 
we write 
$M^{\mu\nu\lambda}_{abc}(k_1,k_2)$ 
as $M^{\mu\nu\lambda}(k_1,k_2)$.

To perform the collinear expansion, as usual, in the ``hadron frame''~\cite{EKT07}
with $q^\mu = (q^0, \vec{q})=(0,0,0,-Q)$,
where both the virtual photon and the initial nucleon 
move along the $z$-axis,
we decompose the relevant gluon momenta $k_i$ ($i=1,2$) as
$k_i^\mu=(k_i\cdot n)p^\mu + (k_i\cdot p)n^\mu + k_\perp^\mu
\equiv x_ip^\mu +\omega^\mu_{\ \,\nu}k_i^\nu$,
with
$\omega^\mu_{\ \,\nu}\equiv g^\mu_{\ \,\nu}-p^\mu n_\nu$: 
Since $p^\mu \sim g^{\mu}_+ Q$ in the hadron frame 
and thus the component 
along $p^\mu$ 
gives the 
leading contribution 
with respect to the hard scale $Q$,
we expand $S_{\mu\nu\lambda}(k_1,k_2)$ around $k_i=x_ip$.  
Expressing also the gluon field $A^\alpha$ 
as
$A^{\alpha}=\left(p^\alpha n_\kappa + \omega^\alpha_{\ \,\kappa}\right) A^\kappa
=p^\alpha n\cdot A + \omega^\alpha_{\ \,\kappa} A^\kappa$,
we note that, in the matrix element $M^{\mu\nu\lambda}(k_1,k_2)$ of (\ref{Mk1k20}), 
the
components associated with the second term $\omega^\alpha_{\ \,\kappa} A^\kappa$
give rise to the contributions
suppressed by $\sim 1/Q$ or more, compared with the corresponding contribution due to 
the first term, $p^\alpha n\cdot A$
(see, e.g., \cite{EKT07}).
We organize the integrand of (\ref{wtensor})
according to the order counting based on those decompositions,
keeping the terms
necessary in the twist-3 accuracy.
For this purpose,
we need the Taylor expansion of the hard partonic subprocesses up to the terms
with the third-order derivative, as
$S_{\mu\nu\lambda}(k_1,k_2)
\simeq S_{\mu\nu\lambda}(x_1,x_2)
+\omega^\alpha_{\ \,\kappa}k_1^\kappa
\left. \partial S_{\mu\nu\lambda}(k_1,k_2) / \partial k_1^\alpha\right|_{k_i=x_ip} +\cdots
+(1/6)\omega^\alpha_{\ \,\kappa}k_2^\kappa 
\,\omega^\beta_{\ \,\tau}k_2^\tau \,\omega^\gamma_{\ \,\sigma}k_2^\sigma  
\left.\partial^3 S_{\mu\nu\lambda}(k_1,k_2)/
\partial k_2^\alpha \partial k_2^\beta \partial k_2^\gamma \right|_{k_i=x_ip}$,
where 
$S_{\mu\nu\lambda}(x_1,x_2)\equiv S_{\mu\nu\lambda}(x_1p,x_2p)$,
because the terms with the third derivatives 
could produce the contributions that behave as the same order as the term,
$S_{\alpha\beta\gamma}(x_1,x_2)
\omega^\alpha_{\ \,\kappa} \omega^\beta_{\ \,\tau}
\omega^\gamma_{\ \,\sigma}M^{\kappa \tau \sigma}(k_1,k_2)$,
arising in the collinear expansion of the integrand of (\ref{wtensor}):
If we substitute $S_{\alpha\beta\gamma}(x_1,x_2)
\omega^\alpha_{\ \,\kappa} \omega^\beta_{\ \,\tau}
\omega^\gamma_{\ \,\sigma}M^{\kappa \tau \sigma}(k_1,k_2)$
directly into the integrand of (\ref{wtensor})
and perform the integrals over $k_1$ and $k_2$,
the result
produces the 
contribution, which is associated with 
the hard scattering between the three physical gluons from the nucleon and
behaves as the same order as the formal convolution of $S_{\alpha\beta\gamma}(x_1,x_2)$
with the twist-3 correlation functions in (\ref{MFabc});
this corresponds to a quantity of twist-3.
Thus, our collinear expansion 
produces lots of terms, 
each of which is not gauge invariant, and
it would look hopeless to
reorganize those into a 
form of the convolution
with only the gauge-invariant correlation functions
$O^{\alpha\beta\gamma}(x_1,x_2)$ and 
$N^{\alpha\beta\gamma}(x_1,x_2)$ 
of (\ref{MFabc}) used.

However, as was the case for the pion production 
associated with the twist-3 quark-gluon correlation functions\,\cite{EKT07}, 
great simplification occurs due to
Ward identities satisfied by the 
corresponding partonic hard-scattering function, $S_{\mu\nu\lambda}(k_1,k_2)$.  
To derive the Ward identities, we note that 
the contribution to $S_{\mu\nu\lambda}(k_1,k_2)$ from
each diagram in figure~\ref{fig2} all contains the two delta functions,
$\delta\left((k_2+q-p_c)^2-m_c^2\right)$
and $\delta\left( (p_c+k_1-k_2)^2-m_c^2\right)$,
representing the on-shell conditions associated, respectively, with the final-state cut on the unobserved 
$\bar{c}$-quark line and with the unpinched pole contribution of the barred propagator.
Also, in each diagram in figure~\ref{fig2}, 
the $c$-quark line fragmenting 
into the final-state $D$-meson
is on-shell (see the discussion below (\ref{Wmunu})).
Due to these on-shell conditions for the diagrams in figure~\ref{fig2}
and the similar conditions for their mirror diagrams, 
$S_{\mu\nu\lambda}(k_1,k_2)$ satisfies the 
Ward identities~\cite{bkty},
\begin{equation}
k_1^\mu S_{\mu\nu\lambda}(k_1,k_2)=0, \;\;\;\;\;\;\;\;\;
k_2^\nu S_{\mu\nu\lambda}(k_1,k_2)=0, \;\;\;\;\;\;\;\;\;
(k_2-k_1)^\lambda S_{\mu\nu\lambda}(k_1,k_2)=0,  
\label{Ward}
\end{equation}
which can be used to
reorganize 
our collinear expansion.
Indeed, combined with another relation,
\beq
\left.{\partial S_{\mu\nu \lambda}(k_1,k_2) p^\lambda \over \partial k_{1\perp}^\sigma }\right|_{k_i=x_ip}=
-\left.{\partial S_{\mu\nu \lambda}(k_1,k_2) p^\lambda \over \partial k_{2\perp}^\sigma }\right|_{k_i=x_ip},
\eeq
which can be derived 
by direct inspection of 
the diagrams in figure~\ref{fig2} and their mirror diagrams,
we find~\cite{bkty} that all the gauge-noninvariant terms in our collinear expansion vanish or cancel among themselves,
and the remaining terms can be expressed 
in a gauge-invariant form,
as the factorization formula
in terms of the 
three-gluon correlation functions of twist-3, defined in (\ref{MFabc}),  
\beq
w\left(p,q,p_c\right)
=\int {dx_1 \over x_1}\int{dx_2\over x_2}
\left. 
{\partial S^{abc}_{\mu\nu\lambda}(k_1,k_2,q,p_c)p^\lambda
\over \partial k_2^\sigma}\right|_{k_i=x_ip}
\omega^\mu_{\ \alpha}\,\omega^\nu_{\ \beta}\,\omega^\sigma_{\ \gamma}\,
{\cal M}^{\alpha\beta\gamma}_{F,abc}(x_1,x_2), 
\label{wfinal}
\eeq
up to the higher-order corrections beyond the present accuracy.
Here, we have restored the color indices $a,b,c$ as well as momentum variables $q, p_c$,
which were associated with the hard-scattering function in (\ref{wtensor}).
Thus, the
{\it total twist-3} contribution to
$w(p,q,p_c)$, relevant to SSA, proves to be expressed solely in terms of the 
three-gluon 
functions $O(x_1,x_2)$ and 
$N(x_1,x_2)$ (see~(\ref{3gluonO})).

When calculating the hard part in (\ref{wfinal}),
$\left.\partial S^{abc}_{\mu\nu\lambda}(k_1,k_2,q,p_c)p^\lambda
/ \partial k_2^\sigma\right|_{k_i=x_ip}$, 
the derivative with respect to $k_2^\sigma$ can hit the delta functions,
$\delta\left((k_2+q-p_c)^2-m_c^2\right)$
and $\delta\left( (p_c+k_1-k_2)^2-m_c^2\right)$, which are 
involved in $S^{abc}_{\mu\nu\lambda}(k_1,k_2,q,p_c)p^\lambda$ as mentioned 
above (\ref{Ward}).
Such derivatives of these delta functions can be 
reexpressed by the derivatives with respect to $x_2$, and then be
treated by integration by parts,
giving rise to the derivative of the three-gluon functions $O(x_1,x_2)$ and 
$N(x_1,x_2)$.
After such manipulations, 
we have
$\left. \delta\left( (p_c+k_1-k_2)^2-m_c^2\right)\right|_{k_i=x_ip} = (1/2p_c\cdot~p) \delta(x_1-~x_2)$,
manifesting the SGP contribution,
while another delta function 
becomes ($\xhat\equiv x_{bj}/ x$, $\zhat\equiv z_f/ z$,
$q_T \equiv P_{h\perp}/z_f$
with  
$x_{bj}=Q^2/ (2p\cdot q)$ and
$z_f=p\cdot P_h/p\cdot q$, as usual)
\beq
\left.\delta\left( (k_2+q-p_c)^2-m_c^2\right)\right|_{k_2=xp} ={1\over \zhat Q^2}
\delta\left( {q_T^2\over Q^2} -\left({1\over \xhat}-1\right)\left({1\over \zhat}-1\right)
+{m_c^2\over \zhat^2 Q^2}\right), 
\label{deltadelta}
\eeq
implying $x >x_{bj}$;
the similar contributions arise also from the corresponding mirror diagrams.
Therefore, evaluating the SGP contributions in the hard part in (\ref{wfinal}),
the resulting SSA
proves to
receive contributions associated with $O(x,x)$, $O(x,0)$, $N(x,x)$, $N(x,0)$,
and also their derivatives with respect to $x$.
In particular, 
the partonic hard part
associated with $O(x,x)$ ($dO(x,x)/dx$) is different from the hard part 
associated with $O(x,0)$ ($dO(x,0)/dx$),
due to the difference
in the tensor structures among 
the three terms in the RHS of (\ref{3gluonO}).
Similarly, the partonic hard part for 
$N(x,x)$ ($dN(x,x)/dx$) is different from that for
$N(x,0)$ ($dN(x,0)/dx$).

Here, we make a
brief comment on the calculation presented in \cite{KQ08} 
for the same phenomenon,
i.e., 
for the twist-3 mechanism to the SSA in SIDIS,
$ep^\uparrow\to eDX$. 
The calculation by Kang and Qiu in \cite{KQ08} started from a factorization formula that was
proposed as a straightforward extension
of the corresponding factorization formula
for the SSA in the pion production, 
associated with the twist-3 quark-gluon correlation functions,
but Kang-Qiu's factorization formula
did not manifest gauge invariance, nor permutation symmetry among the participating gluons
(see equations~(26) and (27) in \cite{KQ08}).
Assuming gauge invariance in their factorization formula, 
Kang and Qiu claimed that
$w\left(p,q,p_c\right)$ can be eventually calculated
with the following formula, 
\beq
\int {dx_1 \over x_1}\int{dx_2\over x_2}
\left. 
{\partial S^{abc}_{\mu\nu\lambda}(k_1,k_2,q,p_c)p^\lambda g_{\perp}^{\mu\nu}
\over \partial k_{2\perp}^\sigma}\right|_{k_i=x_ip}
\omega^\sigma_{\ \gamma}\,
g_{\alpha\beta}
{\cal M}^{\alpha\beta\gamma}_{F,abc}(x_1,x_2),  
\label{KQ}
\eeq
in the notation used in this report,
where 
$g_\perp^{\mu\nu}=g^{\mu\nu}-p^\mu n^\nu -p^\nu n^\mu
=-S^\mu S^\nu - \epsilon^{\mu pnS} \epsilon^{\nu pnS}$, and
we can make the replacement $\omega^\sigma_{\ \gamma} \rightarrow g_{\perp \gamma}^\sigma$,
up to the irrelevant corrections of twist-4 and higher.
Combined with the property,
$S_\gamma g_{\alpha\beta} {\cal M}^{\alpha\beta\gamma}_{F,abc}(x_1,x_2)=0$,
which follows from (\ref{MFabc}), (\ref{3gluonO}),
we see that the three-gluon correlation functions involved in (\ref{KQ})
are expressed by the two types of functions of $x$, $T_G^{(\pm)}(x,x)$, and their derivative,
after evaluating the SGP at $x_1=x_2\equiv x$, where
\beq
T_G^{(\pm)}(x,x)=\int{dy_1^-dy_2^-\over 2\pi}
e^{ixp^+y_1^-} {1\over xp^+} g_{\beta\alpha}\epsilon_{S\gamma n p}\la pS| 
C^{bca}_\pm F_{b}^{\beta +}(0)F_{c}^{\gamma +}(y_2^-)F_a^{\alpha +}(y_1^-)
|pS\ra, 
\label{3gluonKQ}
\eeq
with 
$C_+^{bca}=if^{bca}$, 
$C_-^{bca}= d^{bca}$;
in contrast to our result 
mentioned above,
the contributions $T_G^{(\pm)}(x,0)$, $dT_G^{(\pm)}(x,0)/dx$, associated with unequal arguments, 
do not arise.
The functions~(\ref{3gluonKQ}) 
are given by
the contraction of 
(\ref{MFabc})
with the particular tensor $g_{\beta\alpha}\epsilon_{S\gamma n p}$.
Using (\ref{3gluonO}), we have
\begin{equation}
{xg\over 2\pi} T_G^{(+)}(x,x) = -4M_N\left[N(x,x) -N(x,0) \right], \;\;\;
{xg\over 2\pi} T_G^{(-)}(x,x) = -4M_N\left[O(x,x) +O(x,0) \right].
\end{equation}
Thus,
the 
twist-3 SSA obtained in \cite{KQ08} 
implies the same 
partonic hard parts for $O(x,x)$ and $O(x,0)$ (for 
$dO(x,x)/dx$ and $dO(x,0)/dx$), and similarly for $N(x,x)$ and $-N(x,0)$
(for 
$dN(x,x)/dx$ and $-dN(x,0)/dx$);
clearly, such result contradicts with
the above-mentioned result based on our complete formula~(\ref{wfinal}).  
It is straightforward to see that, 
if 
the tensor structure of the three-gluon correlation function 
${\cal M}^{\alpha\beta\gamma}_{F,abc}(x_1,x_2)$ of (\ref{MFabc})
were assumed to be given by only one structure, $g^{\alpha \beta}\epsilon^{\gamma pnS}$,
our formula~(\ref{wfinal}) would reduce to the formula~(\ref{KQ}),
up to the corrections of twist-4 and higher.
However, such assumption contradicts with the permutation symmetry
required by the Bose statistics of the gluon,
as 
discussed in section 2 and 
represented in (\ref{3gluonO}).

\section{Result for the twist-3 single-spin-dependent cross section for $ep^\uparrow\to eDX$}
We calculate the hadronic tensor~(\ref{Wmunu})
using our factorization formula (\ref{wfinal}),
and evaluate the contraction $L^{\mu\nu}W_{\mu\nu}$
with the leptonic tensor for the unpolarized electron,
$L_{\mu\nu}=2(\ell_\mu\ell_\nu^{\prime}+\ell_\nu\ell_\mu^{\prime})
-Q^2g_{\mu\nu}$.
The result is decomposed according to the dependence on the azimuthal angles
$\phi$ and $\Phi_S$
for the initial-lepton's 3-momentum $\vec{\ell}$ and the nucleon's spin vector $\vec{S}$, respectively,
measured from the axis along the transverse-momentum  $\vec{P}_{h\perp}$  in the hadron frame,
and, through standard manipulations,
leads to 
the spin-dependent, differential cross section
for $ep^\uparrow\to eDX$~\cite{bkty}:
\beq
&&
\hspace{-0.7cm}
\frac{d^5\Delta\sigma
}{[d\omega]}
=\frac{\alpha_{em}^2\alpha_se_c^2 M_N}{8\pi
 x_{bj}^2S_{ep}^2Q^2}\! \left(\frac{-\pi}{2}\right) \!
\sum_{k}
{\cal
 A}_k{\cal S}_k 
\int
\frac{dx}{x}\!
\int
\frac{dz}{z} \delta \! \left(
\frac{q_T^2}{Q^2}-\! \left(1-\frac{1}{\hat{x}}\right) \! \left(1-\frac{1}{\hat{z}}\right)\!
+\frac{m_c^2}{\hat{z}^2Q^2}\right)\!\!\sum_{a=c,\bar{c}}
D_a(z) \nonumber\\
&&
\!\!\!\!\!\!\!\!
\times\! \left(\!\delta_a \!\left\{
\left[\frac{dO(x,x)}{dx}-\frac{2O(x,x)}{x}\right]\! \Delta\hat{\sigma}^{1}_k
+\!\left[\frac{dO(x,0)}{dx}-\frac{2O(x,0)}{x}\right]\! \Delta\hat{\sigma}_k^2
+\!\frac{O(x,x)}{x}\Delta\hat{\sigma}^{3}_k
+\!\frac{O(x,0)}{x}\Delta\hat{\sigma}^{4}_k
\right\}\right. \nonumber\\
&&\!\!\!\!\!\!\!\!\!\!
\left.
+
\! \left[\frac{dN(x,x)}{dx}-\frac{2N(x,x)}{x}\right]\! \Delta\hat{\sigma}^{1}_k
-\! \left[\frac{dN(x,0)}{dx}-\frac{2N(x,0)}{x}\right]\! \Delta\hat{\sigma}^{2}_k
+\!\frac{N(x,x)}{x}\Delta\hat{\sigma}^{3}_k
-\!\frac{N(x,0)}{x}\Delta\hat{\sigma}^{4}_k
\right)\! ,
\label{3gluonresult}
\eeq
where $S_{ep}=(p+\ell)^2$,
we use the shorthand notation, $[d\omega] \equiv dx_{bj}dQ^2dz_fdq_T^2d\phi$, 
for the differential elements,
and the summation $\sum_k$ implies that the subscript $k$ runs over $1,2,3,4,8,9$, with
\beq
&&{\cal A}_1=1+\cosh^2\psi,\;\;\;\;\;\;\;\;
{\cal A}_2=-2,\;\;\;\;\;\;\;\;
{\cal A}_3=-\cos\phi\sinh 2\psi, \;\;\;\;\;\;\;\;
{\cal A}_4=\cos 2\phi\sinh^2\psi,
\nonumber\\
&& {\cal A}_8=-\sin\phi\sinh 2\psi, \;\;\;\;\;\;\;\;
{\cal A}_9=\sin 2\phi\sinh^2\psi , \;\;\;\;\;\;\;\;
(\cosh\psi \equiv {2x_{bj}S_{ep}\over Q^2} -1), 
\label{Ak}
\eeq
and ${\cal S}_k$ defined as
${\cal S}_k=\sin\Phi_S$ for $k=1,2,3,4$
and ${\cal S}_k=\cos\Phi_S$ for $k=8,9$.
The quark-flavor index $a$ can, in principle, be $c$ and $\bar{c}$, with $\delta_c=1$ and  
$\delta_{\bar{c}}=-1$, so that the cross section for 
the $\bar{D}$-meson production $ep^\uparrow\to e\bar{D}X$
can be obtained by 
a simple replacement of the fragmentation function
to that for the $\bar{D}$ meson, $D_a (z) \rightarrow \bar{D}_a (z)$. 
The delta function of (\ref{deltadelta}) appears,
$\alpha_{em}$
is the fine-structure constant,
$\alpha_s=g^2/(4\pi)$ is the strong coupling constant, and
$e_c=2/3$ is the electric charge of the $c$-quark.  
Partonic hard parts $\Delta\hat{\sigma}_k^i$ 
depend on $m_c$ as well as other partonic variables;
for the explicit formulae of $\Delta\hat{\sigma}_k^i$, we refer the readers to \cite{bkty}.
We find that $\Delta\hat{\sigma}_k^1 \neq \Delta\hat{\sigma}_k^2$, $\Delta\hat{\sigma}_k^3 \neq \Delta\hat{\sigma}_k^4$,
while the hard parts 
arising in the second and third lines
in (\ref{3gluonresult})
are common.

The single-spin-dependent cross section (\ref{3gluonresult}) 
can be decomposed into the five structure functions,
based on the different dependences on the azimuthal angles $\Phi_S$ and $\phi$ 
through the above-mentioned explicit forms of ${\cal A}_k$ and ${\cal S}_k$.
The five independent azimuthal structures of this type have been observed also in
the twist-3 single-spin-dependent cross section 
for $ep^\uparrow\to e\pi X$, generated from the quark-gluon correlation functions,
as presented in \cite{KT071,KT09}.  
Introducing the azimuthal angles $\phi_h$ and $\phi_S$ of the hadron plane 
and 
the nucleon's spin vector $\vec{S}$, respectively, as measured from the {\it lepton plane},
they are connected to the above $\Phi_S$ and $\phi$ as 
$\Phi_S=\phi_h-\phi_S$, $\phi=\phi_h$, and
one may express (\ref{3gluonresult}) as the superposition of five sine 
modulations,
\begin{equation}
\frac{d^5\Delta \sigma}{[d\omega]}
=f_1\sin(\phi_h-\phi_S)
+f_2\sin(2\phi_h-\phi_S)
+f_3\sin\phi_S
+f_4\sin(3\phi_h-\phi_S)
+f_5\sin(\phi_h+\phi_S) ,
\label{azimuth2}
\end{equation}
with the corresponding structure functions $f_1,f_2,\ldots, f_5$.
A similar form of five independent azimuthal dependences in the SSA
was obtained in the TMD approach\,\cite{Bacchetta:2006tn}
for small-$P_{h\perp}$ regions.    

It is worth comparing (\ref{3gluonresult}) 
with the 
unpolarized cross section for SIDIS, $ep\to eDX$,
which can be obtained straightforwardly by applying the above-developed formalism 
to extract the twist-2 contribution from figure~\ref{fig1}(a),  
generated from
the unpolarized gluon-density distribution~(\ref{gx}), as
\begin{equation}
\frac{d^5\sigma^{\rm unpol}}{[d\omega]}
=\frac{\alpha_{em}^2\alpha_se_c^2}{32\pi
 x_{bj}^2S_{ep}^2Q^2}
\! \sum_{k}{\cal A}_k \!\!
\int \!
\frac{dx}{x} \!
\int \!
\frac{dz}{z}\delta \! \left(\frac{q_T^2}{Q^2}-\!
\left(1-\frac{1}{\hat{x}}\right) \!\! \left(1-\frac{1}{\hat{z}}\right)\!
+\frac{m_c^2}{\hat{z}^2Q^2}\right)
\!\!\!
\sum_{a=c,\bar{c}} \! D_a(z)G(x)\hat{\sigma}^{U}_k ,
\label{unpolresult}
\end{equation}
in the same notation as in (\ref{3gluonresult}), with $\hat{\sigma}^U_8=\hat{\sigma}^U_9=0$.
Comparing the explicit form~\cite{bkty} of the LO partonic hard cross sections $\hat{\sigma}^U_k$ with
the partonic hard parts 
$\Delta\hat{\sigma}_k^1$ in (\ref{3gluonresult}), 
associated with 
the derivatives, $dO(x,x)/dx$ and $dN(x,x)/dx$, of the three-gluon functions,
one finds the following relations between
the partonic hard parts at
the twist-3 level and those at the twist-2 level:
\beq
\Delta\hat{\sigma}_k^1={2q_T\xhat\over Q^2(1-\zhat)} \hat{\sigma}^U_k.
\label{unlike}
\eeq
The other partonic cross sections at the twist-3 level,
$\Delta\hat{\sigma}_k^i$ ($i=2,3,4$) in (\ref{3gluonresult}),
are also related to the partonic hard scattering parts
at the twist-2 level, although, unlike (\ref{unlike}), 
the corresponding relations cannot be manifested by direct comparison between their explicit formulae.
We find that all these remarkable relations are consequences of the ``master formula'',
\begin{equation}
w\left(p,q,p_c\right)
=-i\pi
\int {dx \over x^2}
{\partial {\cal H}_{\mu\nu}^{abc}(xp,q, p_c)
\over \partial p_{c\perp}^\sigma}
\omega^\mu_{\ \alpha}\,\omega^\nu_{\ \beta}\,\omega^\sigma_{\ \gamma}\,
{\cal M}^{\alpha\beta\gamma}_{F,abc}(x,x), 
\label{master}
\end{equation}
which allows us to calculate (\ref{wfinal})
using the
partonic hard part 
${\cal H}_{\mu\nu}^{abc}(xp,q, p_c)$ for the $2\to2$ Born subprocess,
which coincides with the LO contribution to the  hard part in figure~\ref{fig1}(a), up to the color structure.
The proof of (\ref{master})
will be presented elsewhere.
The similar master formula
was derived \cite{KT071,KT072} for 
various processes 
associated with 
twist-3 quark-gluon correlation functions.

\section{Summary}
We have investigated the SSA
for the $D$-meson production in SIDIS, 
generated from 
the twist-3 three-gluon correlation functions for the nucleon.
We first showed
that there are only two independent three-gluon correlation functions 
of twist-3, 
$O(x_1,x_2)$ and $N(x_1,x_2)$, 
which correspond to two possible ways 
to construct color-singlet combination
composed of three active gluons.
Then, we have formulated the method for calculating the
twist-3 single-spin-dependent cross section generated from the three-gluon correlations.
Our formulation is based on a systematic 
analysis of the relevant diagrams in the Feynman gauge and 
gives all the contribution to the cross section at the twist-3 level in the LO in perturbative QCD, 
guaranteeing the gauge invariance of the result.
As a result, the SSA
occurs as the pole contribution of an internal propagator 
in the partonic hard-scattering subprocess, and the corresponding SGP contribution leads to
the SSA expressed in terms of the
four types of gluonic functions of the relevant momentum fraction $x$:
$O(x,x)$, $O(x,0)$, $N(x,x)$ and $N(x,0)$.  
We find that all these four-types of functions and their derivatives with respect to $x$
contribute to the final form of the single-spin-dependent cross section,
generating five independent structures on relevant azimuthal angles.
We also mentioned the master formula that manifests universal structure behind the SGP contributions,
allowing us to relate the 
$3\to 2$ subprocess relevant for
the twist-3 level to the $2\to 2$ subprocess for the twist-2 level.
Our formalism developed for the twist-3 mechanism of the SSA arising from multi-gluon correlations,
and also the above features revealed in $ep^\uparrow\to eDX$, apply to the SSA in the other processes, like
$A_N$ in $p^\uparrow p\to hX$ 
($h=\pi,\ K,\ D$, etc.)\,\cite{KY10}.

\ack
The work of S.Y. is supported by the Grand-in-Aid for Scientific Research
(No.~22.6032) from the Japan Society of Promotion of Science.


\section*{References}
\begin{thebibliography}{9}
\bibitem{To} Qiu J W and Sterman G 1999 {\it Phys. Rev.} D {\bf 59} 014004; Koike Y and Tomita T 2009 {\it Phys. Lett.} B {\bf 675} 181;
Kanazawa K and Koike Y 2010 {\it Phys. Rev.} D {\bf 82} 034009
\bibitem{JQVY06DY} Ji X D, Qiu J W, Vogelsang W and Yuan F 2006 {\it Phys. Rev.} D {\bf 73} 094017
\bibitem{Bacchetta:2006tn} Bacchetta A {\it et al.}
2007
{\it J. High Energy Phys.} JHEP02(2007)093
\bibitem{JQVY06SIDIS} Ji X D, Qiu J W, Vogelsang W and Yuan F 2006 {\it Phys. Lett.} B {\bf 638} 178
\bibitem{EKT07} Eguchi H, Koike Y and Tanaka K 2006 {\it Nucl. Phys.} B {\bf 752} 1; 2007 {\it Nucl. Phys.} B {\bf 763} 193
\bibitem{bkty}
Beppu H, Koike Y, Tanaka K and Yoshida S 2010
  {\it Phys. Rev.} D {\bf 82} 054005
\bibitem{KQ08}
  Kang Z B and Qiu J W 2008
  {\it Phys. Rev.} D {\bf 78} 034005
\bibitem{Ji92} Ji X D 1992 {\it Phys. Lett.} B {\bf 289} 137; see also A.~V. Belitsky {\it et al.} 2001
{\it Phys. Rev.} D {\bf 63} 094012
\bibitem{Braun09}
Braun V M, Manashov A N and Pirnay B 2009 {\it Phys. Rev.} D {\bf 80} 114002
\bibitem{KT071} Koike Y and Tanaka K 2007 {\it Phys. Lett.} B {\bf 646} 232 [2008 {\it Erratum-ibid.} B {\bf 668} 458]
\bibitem{KT09} Koike Y and Tanaka K 2009 {\it AIP Conf. Proc.} {\bf 1149} 475;
arXiv:0907.2797 [hep-ph] 
\bibitem{KT072} Koike Y and Tanaka K 2007 {\it Phys. Rev.} D {\bf 76} 011502 
\bibitem{KY10} Koike Y and Yoshida S, {\it in these proceedings}
\end{thebibliography}
\end{document}